\documentclass[prb,aps,10pt,twocolumn,showpacs,preprintnumbers,amsmath,amssymb,superscriptaddress]{revtex4-1}

\usepackage{graphicx}
\usepackage{dcolumn}
\usepackage{bm}
\usepackage{amsthm}
\usepackage{amsmath}
\usepackage{amssymb}

\begin{document}

\title{Effect of Dzyaloshinski-Moriya interaction on spin-polarized neutron scattering}

\author{Andreas Michels}\email{andreas.michels@uni.lu}
\author{Denis Mettus}
\affiliation{Physics and Materials Science Research Unit, University of Luxembourg, 162A~Avenue de la Fa\"iencerie, L-1511 Luxembourg, Grand Duchy of Luxembourg}
\author{Dirk Honecker}
\affiliation{Institut Laue-Langevin, 71~avenue des Martyrs, CS20156, F-38042 Grenoble Cedex~9, France}
\author{Konstantin L. Metlov}\email{metlov@fti.dn.ua}
\affiliation{Donetsk Institute for Physics and Technology, Rosa Luxembourg Str.~72, Donetsk, 83114, Ukraine}


\begin{abstract}
For magnetic materials containing many lattice imperfections (\textit{e.g.}, nanocrystalline magnets), the relativistic Dzyaloshinski-Moriya (DM) interaction may result in nonuniform spin textures due to the lack of inversion symmetry at interfaces. Within the framework of the continuum theory of micromagnetics, we explore the impact of the DM interaction on the elastic magnetic small-angle neutron scattering (SANS) cross section of bulk ferromagnets. It is shown that the DM interaction gives rise to a polarization-dependent asymmetric term in the spin-flip SANS cross section. Analysis of this feature may provide a means to determine the DM constant.
\end{abstract}

\maketitle

\section{Introduction}

The Dzyaloshinski-Moriya (DM) interaction \cite{dzya58,moriya60} has recently become anew the focus of an intense worldwide research effort in condensed-matter physics. This interaction is due to relativistic spin-orbit coupling, and in low-symmetry crystal structures lacking inversion symmetry it gives rise to antisymmetric magnetic interactions. Particularly well investigated classes of materials are ultrathin film nanostructures and noncentrosymmetric B20 transition-metal compounds (\textit{e.g.}, MnSi, $\mathrm{Fe}_\mathrm{1-x}\mathrm{Co}_\mathrm{x}\mathrm{Si}$, or FeGe), where the DM interaction plays an important role for the formation of various kinds of inhomogeneous spin structures such as long-wavelength spirals, vortex states, and skyrmion textures (see, \textit{e.g.}, Refs.~\onlinecite{bogdanov89,bogdanov94,bogdanov2005,rossler2006,binz2006,bode07,yamasaki2007,pflei2009,pflei2010,heinze2011,kanazawa2012,milde2013,fert2013,thiaville2013,kostylev2014,wilson2014,wiesendanger2015,rybakov2015} and references therein).

However, already in 1963, Arrott \cite{arrott1963} pointed out that even in a high-symmetry lattice, where the antisymmetric DM term would normally vanish, this interaction is present in the vicinity of any lattice defect. Arrott argued that in antiferromagnetic crystals the DM interaction results in parasitic ferromagnetism, whereas in ferromagnets it gives rise to local antiferromagnetism (in this way reducing the spontaneous magnetization). Hence, based on these considerations, one may expect that the DM interaction substantially influences the magnetic microstructure of polycrystalline materials with a large defect density. In a sense, microstructural defects act as a source of additional local chiral interactions, similar to the above mentioned (intrinsic) DM interactions in noncentrosymmetric crystals. 

One class of materials, where defects play a decisive role, are nanocrystalline magnets, which are characterized by an extremely large interface-to-volume ratio; note that the volume fraction of internal interfaces (\textit{e.g.}, grain boundaries) scales as $L^{-1}$, where $L \sim 10-20 \, \mathrm{nm}$ represents the average crystallite size. This implies that a significant amount of atoms ($\sim 10-20 \, \mathrm{vol.}\%$) in such magnets are localized in the near-vicinity of interfaces, where inversion symmetry is likely to be broken. Consequently, the magnetic properties of a polycrystalline magnet may be substantially influenced by the DM term once its average grain size ``goes nano''.

Thus, the DM interaction might reveal itself in magnets with many crystalline imperfections. Let us now address the question of how to measure an ensuing ``effect''. Traditionally, the influence of lattice defects on the magnetization of bulk magnetic materials is studied by analyzing the high-field branch of a hysteresis curve (see, \textit{e.g.}, the classic studies by Brown and Kronm\"uller). \cite{brown,aharonibook,kronfahn03} However, this approach suffers from the disadvantages that it provides only integral (and no spatially-resolved) information and that the result of such an analysis may depend on the range of applied-field values over which the magnetization data are analyzed. As we will see below, neutron scattering and, in particular, polarized small-angle neutron scattering (SANS) \cite{moon69,michels2010epjb,uehland2010,laver2010,michels2012prb2,kryckaprl2014,niketic2015} provides a unique means to investigate the relevance of the defect-induced DM interaction on a microscopic scale and inside the bulk of inhomogeneous magnets.

The central subject of this paper is the impact of the DM interaction on the elastic magnetic SANS cross sections of lattice-defect-rich bulk ferromagnets. We would like to particularly emphasize that our interest is not directed towards the study of the skyrmion lattice, which is typically investigated by means of neutron diffraction at small scattering angles, \cite{pflei2009} but to the microstructural-defect-induced impact of the DM interaction on the diffuse magnetic neutron scattering at small momentum transfers ($\mathbf{q} \cong 0$). We take advantage of our recently developed micromagnetic SANS theory, \cite{michels2013,metmi2015,mettus2015,metmi2016} which considers magnetic small-angle scattering due to spatially fluctuating saturation magnetization and magnetic anisotropy fields. In addition to the standard magnetic energy contributions such as magnetostatics, magnetic anisotropy, and exchange (see below), we now consider also a phenomenological DM energy term \cite{bogdanov89,bogdanov94,bogdanov2005}
\begin{equation}
E_{DM} = \frac{D}{M_s^2} \int_V \mathbf{M} \cdot (\nabla \times \mathbf{M}) \, dV ,
\label{edmdef}
\end{equation}
where $D$ (in units of $\mathrm{J}/\mathrm{m}^2$) denotes an effective DM constant taking on positive or negative values depending on the material; $\mathbf{M}$ is the magnetization vector field, subject to the constraint $|\mathbf{M}| = M_s$, where $M_s$ is the saturation magnetization. It is also worth emphasizing that, depending on the crystallographic symmetry, the energy of the DM interactions can be written as a combination of different so-called Lifshitz invariants, \cite{bogdanov94} 
\begin{equation}
L^{(k)}_{ij} = M_i \frac{\partial M_j}{\partial k} - M_j \frac{\partial M_i}{\partial k} ,
\label{lifshitz}
\end{equation}
where the $i, j ,k$ are suitable combinations of the Cartesian coordinates $x, y, z$. The particular expression Eq.~(\ref{edmdef}) with
\begin{equation}
L = \mathbf{M} \cdot (\nabla \times \mathbf{M}) = L^{(x)}_{zy} + L^{(y)}_{xz} + L^{(z)}_{yx}
\label{lifshitz_special}
\end{equation}
describes systems having \textit{cubic} symmetry (\textit{e.g.}, MnSi and other B20 compounds). As is shown below, inclusion of Eq.~(\ref{edmdef}) into the micromagnetic energy functional results in a polarization-dependent contribution to the spin-flip SANS cross section. We note that such an asymmetry has also been reported for the spin-wave spectra of magnetic thin films. \cite{Cortes-Ortuno2013}

\section{Micromagnetic theory}

The static equations of micromagnetics for the bulk can be conveniently written as \cite{brown,aharonibook,kronfahn03}
\begin{equation}
\label{torque}
\mathbf{M}(\mathbf{r}) \times \mathbf{H}_{\mathrm{eff}}(\mathbf{r}) = 0 .
\end{equation}
Equations~(\ref{torque}) express the fact that at static equilibrium the torque on the magnetization $\mathbf{M}(\mathbf{r})$ due to an effective magnetic field $\mathbf{H}_{\mathrm{eff}}(\mathbf{r})$ vanishes everywhere inside the material. The effective field is defined as the functional derivative of the ferromagnet's total energy-density functional $\epsilon$ with respect to the magnetization,
\begin{equation}
\label{heff}
\mathbf{H}_{\mathrm{eff}} = - \frac{1}{\mu_0} \frac{\delta \epsilon}{\delta \mathbf{M}} = \mathbf{H}_0 + \mathbf{H}_d + \mathbf{H}_p + \mathbf{H}_{ex} + \mathbf{H}_{DM} ;
\end{equation}
it is composed of a uniform applied magnetic field $\mathbf{H}_0$, the magnetostatic field $\mathbf{H}_d(\mathbf{r})$, the magnetic anisotropy field $\mathbf{H}_p(\mathbf{r})$, the exchange field $\mathbf{H}_{ex} = l_M^2 \Delta \mathbf{M}$, and of the field $\mathbf{H}_{DM} = - l_D \nabla \times \mathbf{M}$, which is due to the DM interaction; $\Delta$ is the Laplace operator and $\nabla = \partial/\partial x \, \mathbf{e}_x + \partial/\partial y \, \mathbf{e}_y + \partial/\partial z \, \mathbf{e}_z$ is the gradient operator, where the unit vectors along the Cartesian laboratory axes are, respectively, denoted with $\mathbf{e}_x$, $\mathbf{e}_y$, and $\mathbf{e}_z$ ($\mu_0$: vacuum permeability). The micromagnetic length scales
\begin{equation}
\label{lmmmdef}
l_M = \sqrt{\frac{2 A}{\mu_0 M_s^2}}
\end{equation}
and
\begin{equation}
\label{lddddef}
l_D = \frac{2 D}{\mu_0 M_s^2}
\end{equation}
are, respectively, related to the magnetostatic and to the DM interaction. In the present work, the values for the DM constant $D$ and for the exchange-stiffness constant $A$ are assumed to be uniform throughout the material (in contrast to the local saturation magnetization, see below). Using the materials parameters of Table~\ref{table1}, one finds $l_M \cong l_D \cong 5 \, \mathrm{nm}$.

\begin{table}
\caption{\label{table1} Magnetic and structural parameters used for displaying the Fourier components and the SANS cross sections.}
\begin{ruledtabular}
\begin{tabular}{lccccc}
$\mu_0 M_s$~($\mathrm{T}$) & 1.0 \\
\hline
$A$~($\mathrm{pJ/m}$) & 10 \\
\hline
$D$~($\mathrm{mJ/m^2}$) & 2 \\
\hline
$H_p / \Delta M$ & 1 \\
\hline
$R$~(nm) & 8 \\
\end{tabular}
\end{ruledtabular}
\end{table}

In our analysis, we assume the material to be nearly saturated by a strong applied magnetic field $\mathbf{H}_0 \parallel \mathbf{e}_z$, and we express the magnetization as
\begin{equation}
\label{mvecdef}
\mathbf{M}(\mathbf{r}) = \left\{ M_x(\mathbf{r}), M_y(\mathbf{r}), M_s(\mathbf{r}) \right\} ,
\end{equation}
where $M_x \ll M_s$ and $M_y \ll M_s$ (small-misalignment approximation). The local saturation magnetization is assumed to be a function of the position $\mathbf{r} = \left\{ x, y, z \right\}$ inside the material: \cite{schloemann67,metmi2015}
\begin{equation}
\label{msatdef}
M_s(\mathbf{r}) = M_s [1 + I_m(\mathbf{r})] ,
\end{equation}
where $I_m$ is an inhomogeneity function, small in magnitude, which describes the local variation of $M_s$. The spatial average of $I_m$ vanishes, $\langle I_m(\mathbf{r}) \rangle = 0$, so that $\langle M_s(\mathbf{r}) \rangle = M_s$ is the saturation magnetization (which can be measured with a magnetometer). Note that due to the constraint $\left| \mathbf{M} \right| = M_s$, there are only two independent components of $\mathbf{M}$. By defining the Fourier transform $\widetilde{F}(\mathbf{q})$ of a continuous function $F(\mathbf{r})$ as
\begin{equation}
\label{fourierdef}
\widetilde{F}(\mathbf{q}) = \frac{1}{(2\pi)^{3/2}} \int_V F(\mathbf{r}) \, \exp(-i \mathbf{q} \mathbf{r}) \, dV ,
\end{equation}
where $i^2 = -1$ and $\mathbf{q} = \left\{ q_x, q_y, q_z \right\}$ is the wavevector, one can write for the magnetization Fourier coefficient $\widetilde{\mathbf{M}}(\mathbf{q})$ up to the first order in $\widetilde{I}_m$
\begin{equation}
\label{mvecdeffourier}
\widetilde{\mathbf{M}}(\mathbf{q}) = \left\{ \widetilde{M}_x(\mathbf{q}), \widetilde{M}_y(\mathbf{q}), M_s [\delta(\mathbf{q}) + \widetilde{I}_m(\mathbf{q})] \right\} ,
\end{equation}
where $\delta(\mathbf{q})$ is the Dirac's Delta function. 

The remaining fields in the balance-of-torques Eq.~(\ref{torque}) also have convenient analytical representations in terms of their Fourier transforms. The magnetostatic field $\mathbf{H}_d(\mathbf{r})$ can be written as the sum of the demagnetizing field $\mathbf{H}_d^s(\mathbf{r})$ due to surface charges and of the magnetostatic field $\mathbf{H}_d^b(\mathbf{r})$ which is related to volume charges, \textit{i.e.}, $\mathbf{H}_d(\mathbf{r}) = \mathbf{H}_d^s(\mathbf{r}) + \mathbf{H}_d^b(\mathbf{r})$. In the high-field limit (when the magnetization is close to saturation) and for samples with an ellipsoidal shape with $\mathbf{H}_0$ directed along a principal axis of the ellipsoid, one may approximate the demagnetizing field due to the surface charges by the \textit{uniform} field $\mathbf{H}_d^s = - N_d M_s \mathbf{e}_z$, where $0 < N_d < 1$ denotes the corresponding demagnetizing factor. The field $H_d^s = - N_d M_s$ can then be combined with the applied magnetic field $H_0$ to yield the internal magnetic field $H_i = H_0 - N_d M_s$ [\textit{i.e.}, the $\mathbf{q} = 0$ Fourier component of $\mathbf{H}_d(\mathbf{r})$]. \cite{aharonibook} At $\mathbf{q} \neq 0$, the Fourier component $\widetilde{\mathbf{H}}_d^b(\mathbf{q})$ of $\mathbf{H}_d^b(\mathbf{r})$ is found from basic magnetostatics ($\nabla \cdot \mathbf{B} = 0$; $\nabla \times \mathbf{H}_d = 0$): \cite{herring51} 
\begin{equation}
\label{hdbbbdef}
\widetilde{\mathbf{H}}_d^b(\mathbf{q}) = - \frac{\mathbf{q} [\mathbf{q} \cdot \mathbf{\widetilde{M}}(\mathbf{q})]}{q^2} ,
\end{equation}
so that the total $\widetilde{\mathbf{H}}_d(\mathbf{q})$ is
\begin{equation}
\label{hdbbbdeftot}
\widetilde{\mathbf{H}}_d(\mathbf{q}) = H_i \delta(\mathbf{q}) \mathbf{e}_z - \frac{\mathbf{q} [\mathbf{q} \cdot \mathbf{\widetilde{M}}(\mathbf{q})]}{q^2} .
\end{equation}
The magnetic anisotropy field $\mathbf{H}_p(\mathbf{r})$, with its Fourier transform
\begin{equation}
\label{hpfourierdef}
\widetilde{\mathbf{H}}_p(\mathbf{q}) = \left\{ \widetilde{H}_{p,x}(\mathbf{q}), \widetilde{H}_{p,y}(\mathbf{q}), 0 \right\} ,
\end{equation} 
is a source of spin disorder, since it increases the magnitude of the transversal Fourier components (see below). The field $\mathbf{H}_p(\mathbf{r})$ contains the information about the sample's microstructure (\textit{e.g.}, crystallite size, inhomogeneous lattice strain, crystallographic texture). \cite{jprb2001} Note that no assumption is made about the particular form of the magnetic anisotropy (magnetocrystalline and/or magnetoelastic). We also remind the reader that due to $\left| \mathbf{M} \right| = M_s$, there are only two independent components of $\mathbf{H}_p(\mathbf{r})$. The Fourier representations of the exchange field $\mathbf{H}_{ex}$ and of the DM field $\mathbf{H}_{DM}$ read, respectively,
\begin{equation}
\label{hexfourierdef}
\widetilde{\mathbf{H}}_{ex}(\mathbf{q}) = - l_M^2 q^2 \left\{ \widetilde{M}_x, \widetilde{M}_y, M_s \widetilde{I}_m \right\} ,
\end{equation} 
and
\begin{eqnarray}
\label{hdmfourierdef}
\widetilde{\mathbf{H}}_{DM}(\mathbf{q}) = -i l_D \left\{ q_y M_s \widetilde{I}_m - q_z \widetilde{M}_y, \right. \nonumber \\ \left. q_z \widetilde{M}_x - q_x M_s \widetilde{I}_m, q_x \widetilde{M}_y - q_y \widetilde{M}_x \right\} .
\end{eqnarray}
In case of a weak spatial dependency of $A$ and $D$ [with fluctuations of the order of the saturation-magnetization fluctuation $I_m(\mathbf{r})$] the quantities $l_M$ and $l_D$ must be understood as spatial averages of the corresponding (now position-dependent) expressions~(\ref{lmmmdef})$-$(\ref{lddddef}). In this situation, the first-order expansion of the Brown's equations~(\ref{torque}) (see below) is still valid and contains no additional terms. \cite{metmi2015}

By using Eqs.~(\ref{hdbbbdeftot})$-$(\ref{hdmfourierdef}) in the balance of torque equation and by neglecting terms of higher than linear order in $\widetilde{M}_x$, $\widetilde{M}_y$, and $\widetilde{I}_m$ (including terms such as $\widetilde{M}_x \widetilde{I}_m$ and $\widetilde{H}_{p,x} \widetilde{I}_m$), we obtain, in Fourier space, and for a general orientation of the wavevector $\mathbf{q} = \left\{ q_x, q_y, q_z \right\}$, the following set of linear equations for $\widetilde{M}_x$ and $\widetilde{M}_y$: \cite{aharonibook}
\begin{equation}
\label{solmxintermediate}
\widetilde{M}_x \left( 1 + p \frac{q_x^2}{q^2} \right) + \widetilde{M}_y \left( p \frac{q_x q_y}{q^2} - i p \, l_D q_z \right) = K_1 ,
\end{equation} 
\begin{equation}
\label{solmyintermediate}
\widetilde{M}_y \left( 1 + p \frac{q_y^2}{q^2} \right) + \widetilde{M}_x \left( p \frac{q_x q_y}{q^2} + i p \, l_D q_z \right) = K_2 ,
\end{equation} 
where
\begin{equation}
\label{Adef}
K_1 = p \left( \widetilde{H}_{p,x} - M_s \widetilde{I}_m \left[ \frac{q_x q_z}{q^2} + i l_D q_y \right] \right) , 
\end{equation}
and
\begin{equation}
\label{Bdef}
K_2 = p \left( \widetilde{H}_{p,y} - M_s \widetilde{I}_m \left[ \frac{q_y q_z}{q^2} - i l_D q_x \right] \right) .
\end{equation}
The dimensionless function
\begin{equation}
\label{pppdef}
p(q, H_i) = \frac{M_s}{H_{\mathrm{eff}}(q, H_i)}
\end{equation} 
depends on the effective magnetic field
\begin{equation}
\label{heffdef}
H_{\mathrm{eff}} = \left( 1 + l_H^2 q^2 \right) H_i, 
\end{equation} 
and
\begin{equation}
\label{lhdef}
l_H(H_i) = \sqrt{\frac{2 A}{\mu_0 M_s H_i}}
\end{equation} 
denotes the micromagnetic exchange length of the field, which is a measure for the size of inhomogeneously magnetized regions around microstructural lattice defects. 

The present approach is of the first order in the amplitude of the inhomogeneity function $\widetilde{I}_m$, which entails the neglect of complicated convolution products. We refer to Ref.~\onlinecite{metmi2015} for a micromagnetic SANS theory which is up to the second order in $\widetilde{I}_m$.

The solutions of Eqs.~(\ref{solmxintermediate}) and (\ref{solmyintermediate}) are:
\begin{widetext}
\begin{equation}
\label{solmxgeneral}
\widetilde{M}_x = \frac{p \left( \widetilde{H}_{p,x} \left[ 1 + p \frac{q_y^2}{q^2} \right] - M_s \widetilde{I}_m \frac{q_x q_z}{q^2} \left[ 1 + p \, l_D^2 q^2 \right] - \widetilde{H}_{p,y} p \frac{q_x q_y}{q^2}  - i \left[ M_s \widetilde{I}_m (1 + p) l_D q_y - \widetilde{H}_{p,y} p \, l_D q_z \right] \right)}{1 + p \frac{q_x^2 + q_y^2}{q^2} - p^2 l_D^2 q_z^2} , 
\end{equation}
\begin{equation}
\label{solmygeneral}
\widetilde{M}_y = \frac{p \left( \widetilde{H}_{p,y} \left[ 1 + p \frac{q_x^2}{q^2} \right] - M_s \widetilde{I}_m \frac{q_y q_z}{q^2} \left[ 1 + p \, l_D^2 q^2 \right] - \widetilde{H}_{p,x} p \frac{q_x q_y}{q^2}  + i \left[ M_s \widetilde{I}_m (1 + p) l_D q_x - \widetilde{H}_{p,x} p \, l_D q_z \right] \right)}{1 + p \frac{q_x^2 + q_y^2}{q^2} - p^2 l_D^2 q_z^2} ,
\end{equation}
\end{widetext}
Besides computing the magnetic SANS cross section (see below), the above results for the transversal Fourier components [Eqs.~(\ref{solmxgeneral}) and (\ref{solmygeneral})] can also be used for obtaining the autocorrelation function of the spin-misalignment \cite{mettus2015} and the approach-to-saturation law; \cite{kronfahn03} in this way one can investigate the impact of the DM interaction on the high-field magnetization.

\begin{figure}[tb]
\centering
\resizebox{1.0\columnwidth}{!}{\includegraphics{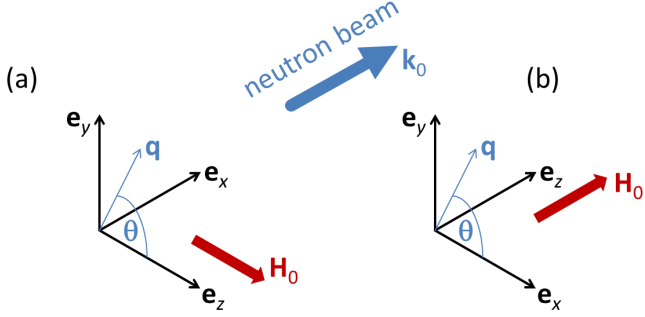}}
\caption{Sketch of the (a) perpendicular and (b) parallel scattering geometry. The applied-field direction $\mathbf{H}_0$ defines in both cases the $\mathbf{e}_z$-direction of a Cartesian laboratory coordinate system. The angle $\theta$ specifies the orientation of the scattering vector $\mathbf{q}$ on the two-dimensional detector: (a) $\mathbf{q} \cong \left\{ 0, q_y, q_z \right\} = q \left\{ 0, \sin\theta, \cos\theta \right\}$ and (b) $\mathbf{q} \cong \left\{ q_x, q_y, 0 \right\} = q \left\{ \cos\theta, \sin\theta, 0 \right\}$.}
\label{fig1}
\end{figure}

\section{Connecting micromagnetics and SANS: Averages of the magnetization Fourier components}

Regarding SANS experiments, two scattering geometries are commonly of relevance: the perpendicular scattering geometry, which has the wavevector $\mathbf{k}_0$ of the incoming neutron beam perpendicular to the applied magnetic field $\mathbf{H}_0 \parallel \mathbf{e}_z$, and the parallel geometry, where $\mathbf{k}_0 \parallel \mathbf{H}_0$ (see Fig.~\ref{fig1}). Furthermore, since SANS probes only correlations in the plane perpendicular to the incident beam, the scattering or momentum-transfer vectors for these two geometries reduce to:
\begin{equation}
\label{qperp}
\mathbf{q} \cong \left\{ 0, q_y, q_z \right\} = q \left\{ 0, \sin\theta, \cos\theta \right\} \hspace{0.5cm} (\mathbf{k}_0 \perp \mathbf{H}_0) ,
\end{equation} 
and
\begin{equation}
\label{qpara}
\mathbf{q} \cong \left\{ q_x, q_y, 0 \right\} = q \left\{ \cos\theta, \sin\theta, 0 \right\} \hspace{0.5cm} (\mathbf{k}_0 \parallel \mathbf{H}_0) .
\end{equation}
\begin{widetext}
For $q_x = 0$ ($\mathbf{k}_0 \perp \mathbf{H}_0$), Eqs.~(\ref{solmxgeneral}) and (\ref{solmygeneral}) reduce to
\begin{equation}
\label{solmxqx0}
\widetilde{M}_x = \frac{p \left( \widetilde{H}_{p,x} \left[ 1 + p \frac{q_y^2}{q^2} \right] - i \left[ M_s \widetilde{I}_m (1 + p) l_D q_y - \widetilde{H}_{p,y} p \, l_D q_z \right] \right)}{1 + p \frac{q_y^2}{q^2} - p^2 l_D^2 q_z^2} , 
\end{equation}
\begin{equation}
\label{solmyqx0}
\widetilde{M}_y = \frac{p \left( \widetilde{H}_{p,y} - M_s \widetilde{I}_m \frac{q_y q_z}{q^2} \left[ 1 + p \, l_D^2 q^2 \right] - i \widetilde{H}_{p,x} p \, l_D q_z \right)}{1 + p \frac{q_y^2}{q^2} - p^2 l_D^2 q_z^2} ,
\end{equation}
whereas for $q_z = 0$ ($\mathbf{k}_0 \parallel \mathbf{H}_0$), we find
\begin{equation}
\label{solmxqz0}
\widetilde{M}_x = \frac{p \left( \widetilde{H}_{p,x} \left[ 1 + p \frac{q_y^2}{q^2} \right] - \widetilde{H}_{p,y} p \frac{q_x q_y}{q^2} - i M_s \widetilde{I}_m (1 + p) l_D q_y \right) }{1 + p} ,
\end{equation}
\begin{equation}
\label{solmyqz0}
\widetilde{M}_y = \frac{p \left( \widetilde{H}_{p,y} \left[ 1 + p \frac{q_x^2}{q^2} \right] - \widetilde{H}_{p,x} p \frac{q_x q_y}{q^2} + i M_s \widetilde{I}_m (1 + p) l_D q_x \right) }{1 + p} .\end{equation}
\end{widetext}

Several comments are in place: (i) We note that both Fourier components $\widetilde{M}_x(\mathbf{q})$ and $\widetilde{M}_y(\mathbf{q})$ are complex functions, which (at $q \neq 0$) depend explicitly on the longitudinal magnetization Fourier coefficient
\begin{equation}
\label{mzzzdef}
\widetilde{M}_z(\mathbf{q}) = M_s \widetilde{I}_m(\mathbf{q}) . 
\end{equation} 
Since $\widetilde{M}_z \propto \Delta M$, \cite{schloemann67} this term models inhomogeneities in the magnetic microstructure that are due to jumps $\Delta M$ in the magnetization at internal interfaces (\textit{e.g.}, particle-matrix boundaries). Furthermore, the magnetic anisotropy field Fourier component $\widetilde{H}_p$ and $\widetilde{M}_z$ both exhibit a tendency to increase the amplitudes of the transversal Fourier coefficients and are, thus, the sources of spin disorder in the system (this is best seen by inspecting the averaged squared functions Eqs.~(\ref{mxp})$-$(\ref{mxmypara}) below). (ii) Terms in $\widetilde{M}_x$ and $\widetilde{M}_y$ such as $q_y^2 / q^2$ or $q_x q_y / q^2$ are due to the long-range magnetodipolar interaction [compare to the above expression for $\widetilde{\mathbf{H}}_d(\mathbf{q})$]. These contributions (in combination with terms related to the DM interaction) give rise to an angular anisotropy \textit{already} in the Fourier components (see Figs.~\ref{fig2} and \ref{fig3}); this anisotropy ($\theta$-dependence, see below) adds on top of the anisotropy that is related to the trigonometric functions originating from the dipolar neutron-magnetic interaction. (iii) The denominator of $\widetilde{M}_x$ and $\widetilde{M}_y$ has (for $\mathbf{k}_0 \perp \mathbf{H}_0$) a singularity for $1 + p q_y^2 / q^2 = p^2 l_D^2 q_z^2$, which becomes particularly noticeable at small fields and for small $q$ along the horizontal direction ($\theta = 0$) where $q_y = 0$. For large $q$ or large $H_i$, the effective magnetic field takes on large values, \cite{michels08rop} so that $p \ll 1$ and the term $p^2 l_D^2 q_z^2$ is much smaller than $1 + p q_y^2 / q^2$. However, we remind that the present theory is valid in the approach-to-saturation regime when the sample consists of a single magnetic domain and one considers small deviations of magnetic moments (due to spatially varying $H_p$, $M_s$, and due to the DM interaction) relative to the applied field direction. (iv) Note the symmetry of the equations in the parallel geometry, which is absent in the perpendicular case. Without the DM interaction ($l_D = 0$), Eqs.~(\ref{solmxqx0})$-$(\ref{solmyqz0}) reduce to Eqs.~(42)$-$(45) in Ref.~\onlinecite{michels2014review}.

Since the magnetic SANS cross sections depend on the magnetization Fourier coefficients (see below), it is necessary to compute appropriate averages of functions such as $|\widetilde{M}_x|^2$, $|\widetilde{M}_y|^2$, $- (\widetilde{M}_y \widetilde{M}_z^{\ast} + \widetilde{M}_y^{\ast} \widetilde{M}_z)$, or $- (\widetilde{M}_x \widetilde{M}_y^{\ast} + \widetilde{M}_x^{\ast} \widetilde{M}_y)$, where the asterisks ``$\, ^{*} \,$'' mark the complex-conjugated quantity (see Ref.~\onlinecite{michels2014review} for a compilation of the various unpolarized and spin-polarized SANS cross sections). For this purpose, we assume that $\widetilde{M}_z = M_s \widetilde{I}_m$ is real-valued ($\widetilde{M}_z = \widetilde{M}_z^{\ast}$) and isotropic [$\widetilde{M}_z = \widetilde{M}_z(q)$], and that the Fourier coefficient $\mathbf{\widetilde{H}}_p(\mathbf{q})$ of the magnetic anisotropy field $\mathbf{H}_p(\mathbf{r})$ is isotropically distributed in the plane perpendicular to $\mathbf{H}_0$, \textit{i.e.},
\begin{equation}
\label{hbetadef}
\mathbf{\widetilde{H}}_p(\mathbf{q}) = \left\{ \widetilde{H}_{p,x}, \widetilde{H}_{p,y}, 0 \right\} = \left\{ \widetilde{H}_p \cos\beta, \widetilde{H}_p \sin\beta, 0 \right\} ,
\end{equation}
where the angle $\beta$ specifies the orientation of $\widetilde{H}_p$; in other words, we assume that the vector $\mathbf{\widetilde{H}}_p(\mathbf{q})$ takes on all orientations (angles $\beta$) with equal probability. This allows us to average $| \widetilde{M}_x(q, \theta, H_i, \beta) |^2$, etc.\ over the angle $\beta$, according to 
\begin{equation}
\label{hbetaaveragedef}
(2\pi)^{-1} \int_0^{2\pi} (...) d\beta \nonumber .
\end{equation}

\begin{widetext}
The results for the perpendicular scattering geometry ($\mathbf{k}_0 \perp \mathbf{H}_0$) are:
\begin{equation}
\label{mxp}
|\widetilde{M}_x|^2 = \frac{p^2}{2} \frac{\widetilde{H}^2_p \left( \left[ 1 + p \sin^2\theta \right]^2 + p^2 l_D^2 q^2 \cos^2\theta \right) + 2 \widetilde{M}_z^2 (1 + p)^2 l_D^2 q^2 \sin^2\theta}{\left( 1 + p \sin^2\theta - p^2 l_D^2 q^2 \cos^2\theta \right)^2} , 
\end{equation}
\begin{equation}
\label{myp}
|\widetilde{M}_y|^2 = \frac{p^2}{2} \frac{\widetilde{H}_p^2 \left( 1 + p^2 l_D^2 q^2 \cos^2\theta \right) + 2 \widetilde{M}_z^2 \left( 1 + p \, l_D^2 q^2 \right)^2 \sin^2\theta \cos^2\theta}{\left( 1 + p \sin^2\theta - p^2 l_D^2 q^2 \cos^2\theta \right)^2} , 
\end{equation}
\begin{equation}
\label{mymzp}
- (\widetilde{M}_y \widetilde{M}_z^{\ast}  + \widetilde{M}_y^{\ast} \widetilde{M}_z) = \frac{2 \widetilde{M}_z^2 p \left( 1 + p \, l_D^2 q^2 \right) \sin\theta \cos\theta}{1 + p \sin^2\theta - p^2 l_D^2 q^2 \cos^2\theta} .
\end{equation}
The results for the parallel scattering geometry ($\mathbf{k}_0 \parallel \mathbf{H}_0$) are:
\begin{equation}
\label{mxpara}
|\widetilde{M}_x|^2 = \frac{p^2}{2} \frac{\widetilde{H}_p^2 \left( 1 + p (2 + p) \sin^2\theta \right) +
2 \widetilde{M}_z^2 (1 + p)^2 l_D^2 q^2 \sin^2\theta}{\left( 1 + p \right)^2} ,
\end{equation}
\begin{equation}
\label{mypara}
|\widetilde{M}_y|^2 = \frac{p^2}{2} \frac{\widetilde{H}_p^2 \left( 1 + p (2 + p) \cos^2\theta \right) + 
2 \widetilde{M}_z^2 (1 + p)^2 l_D^2 q^2 \cos^2\theta}{\left( 1 + p \right)^2} ,
\end{equation}
\begin{equation}
\label{mxmypara}
- (\widetilde{M}_x \widetilde{M}_y^{\ast}  + \widetilde{M}_x^{\ast} \widetilde{M}_y) = p^2 \frac{\left( \widetilde{H}_p^2 p (2 + p) + 2 \widetilde{M}_z^2 (1 + p)^2 l_D^2 q^2 \right) \sin\theta \cos\theta}{\left( 1 + p \right)^2} .
\end{equation}
\end{widetext}

For graphically displaying the Fourier components and SANS cross sections, we employ the sphere form factor for both $\widetilde{H}^2_p$ and $\widetilde{M}^2_z$ (Ref.~\onlinecite{mettus2015}):
\begin{equation}
\label{hsquaredmodel}
\widetilde{H}^2_p(q) = \frac{H_p^2}{(8\pi)^3} P(q) ,
\end{equation}
\begin{equation}
\label{mzsquaredmodel}
\widetilde{M}^2_z(q) = \frac{(\Delta M)^2}{(8\pi)^3} P(q) ,
\end{equation}
\begin{equation}
\label{spheremodel}
P(q) = 9 V_p^2 \frac{j^2_1(q R)}{(q R)^2} ;
\end{equation}
$V_p = \frac{4\pi}{3} R^3$ and $j_1(x)$ denotes the spherical Bessel function of first order. We also note that the characteristic structure sizes of $\widetilde{H}^2_p$ and $\widetilde{M}^2_z$ need not to be identical; these are related, respectively, to the spatial extent of regions with uniform magnetic anisotropy field and saturation magnetization. A corresponding distribution function characterizing $\widetilde{H}^2_p$ and $\widetilde{M}^2_z$ can also be included (see Fig.~\ref{fig5} below). Under the above assumptions, the functions $\widetilde{H}^2_p(q)$ and $\widetilde{M}^2_z(q)$ differ only by constant prefactors, \textit{i.e.}, the magnitude $H_p$ of the mean magnetic anisotropy field and the jump $\Delta M$ of the magnitude of the magnetization at internal interfaces. In addition to $H_p$ and $\Delta M$, we note that the Fourier components depend on the applied magnetic field $H_i$, the magnetic interactions ($A$, $M_s$, $D$), and on the magnitude $q$ and orientation $\theta$ of the scattering vector. The averaged Fourier coefficients, Eqs.~(\ref{mxp})$-$(\ref{mxmypara}), need only to be multiplied by the corresponding trigonometric functions and summed up in order to obtain the magnetic SANS cross section.

\begin{figure}[htb]
\centering
\resizebox{1.0\columnwidth}{!}{\includegraphics{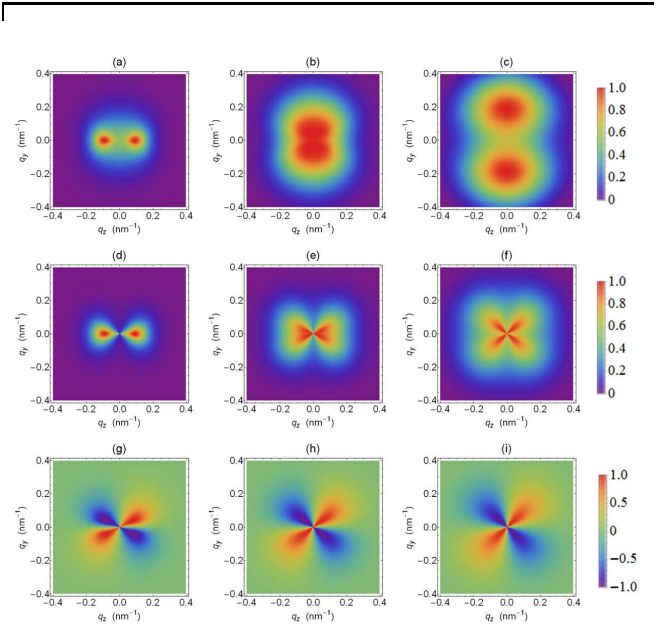}}
\caption{Contour plots of the Fourier components of the magnetization at selected applied magnetic fields ($\mathbf{k}_0 \perp \mathbf{H}_0$). $|\widetilde{M}_x|^2$ (upper row), $|\widetilde{M}_y|^2$ (middle row), and $CT = - (\widetilde{M}_y \widetilde{M}_z^{\ast} + \widetilde{M}_y^{\ast} \widetilde{M}_z)$ (lower row)  [Eqs.~(\ref{mxp})$-$(\ref{mymzp})]. $\mathbf{H}_0 \parallel \mathbf{e}_z$ is horizontal in the plane. $H_i$ values (in T) from left to right column: 0.5; 1.5; 10.0. All data were normalized to unity by the respective maximum value.}
\label{fig2}
\end{figure}
\begin{figure}[htb]
\centering
\resizebox{1.0\columnwidth}{!}{\includegraphics{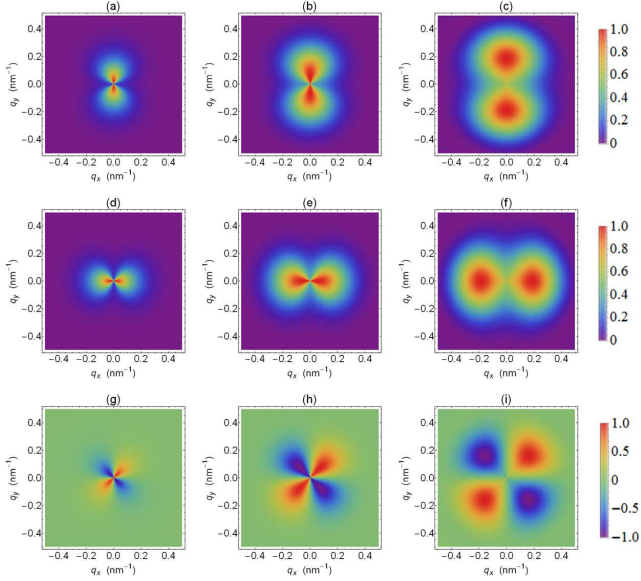}}
\caption{Same as Fig.~\ref{fig2}, but for $\mathbf{k}_0 \parallel \mathbf{H}_0$. $|\widetilde{M}_x|^2$ (upper row), $|\widetilde{M}_y|^2$ (middle row), and $CT = - (\widetilde{M}_x \widetilde{M}_y^{\ast} + \widetilde{M}_x^{\ast} \widetilde{M}_y)$ (lower row)  [Eqs.~(\ref{mxpara})$-$(\ref{mxmypara})].}
\label{fig3}
\end{figure}

Figures~\ref{fig2} and \ref{fig3} visualize, respectively, for $\mathbf{k}_0 \perp \mathbf{H}_0$ and $\mathbf{k}_0 \parallel \mathbf{H}_0$ the angular anisotropy of the Fourier coefficients on a two-dimensional detector; Table~\ref{table1} lists the materials and structural parameters used. It is seen that all Fourier coefficients are highly anisotropic, \textit{e.g.}, $|\widetilde{M}_x|^2$ for $\mathbf{k}_0 \perp \mathbf{H}_0$ changes with increasing field from horizontally to vertically elongated. These anisotropies are clearly a consequence of the magnetodipolar interaction and of terms such as $l_D^2 q^2 \cos^2\theta$ related to the DM interaction. Note that both cross terms ($CT$) change sign at the borders between quadrants [Figs.~\ref{fig2}(g)$-$\ref{fig2}(i) and Figs.~\ref{fig3}(g)$-$\ref{fig3}(i)]. However, on multiplication with $\sin\theta \cos\theta$ (in order to arrive at the corresponding contribution to the cross section, compare Eq.~(\ref{sigmasansperpunpol}) below) these terms become positive definite (at least for not too small $q$ and $H_i$).

\section{Magnetic SANS cross sections}

Although the averages of the magnetization Fourier components for $\mathbf{k}_0 \parallel \mathbf{H}_0$ are highly anisotropic (Fig.~\ref{fig3}), the ensuing magnetic SANS cross sections are isotropic ($\theta$-independent) for statistically isotropic materials. Therefore, we discuss for the remainder of this paper only the (unpolarized and spin-polarized) SANS cross sections for the perpendicular scattering geometry.

\subsection{Unpolarized SANS ($\mathbf{k}_0 \perp \mathbf{H}_0$)}

For $\mathbf{k}_0 \perp \mathbf{H}_0$, the elastic unpolarized SANS cross section $d \Sigma / d \Omega$ at scattering vector $\mathbf{q}$ can be written as: \cite{michels2014review}
\begin{eqnarray}
\label{sigmasansperpunpol}
\frac{d \Sigma}{d \Omega}(\mathbf{q}) = \frac{8 \pi^3}{V} \left( |\widetilde{N}|^2 + b_H^2 |\widetilde{M}_x|^2 + b_H^2 |\widetilde{M}_y|^2 \cos^2\theta \right. \nonumber \\ \left. + b_H^2 |\widetilde{M}_z|^2 \sin^2\theta - b_H^2 (\widetilde{M}_y \widetilde{M}_z^{\ast} + \widetilde{M}_y^{\ast} \widetilde{M}_z) \sin\theta \cos\theta \right) \nonumber ; \\
\end{eqnarray}
$|\mathbf{q}| = q = (4 \pi/ \lambda) \sin \psi$, where $\psi$ is half the scattering angle and $\lambda$ is the wavelength of the incident radiation, $V$ is the scattering volume, $b_H = 2.91 \times 10^{8} \, \mathrm{A^{-1} m^{-1}}$ relates the atomic magnetic moment to the Bohr magneton, $\widetilde{N}(\mathbf{q})$ and $\widetilde{M}_{x, y, z}(\mathbf{q})$ denote, respectively, the Fourier coefficients of the nuclear scattering-length density and of the Cartesian components of the magnetization $\mathbf{M}(\mathbf{r})$, and $\theta$ represents the angle between $\mathbf{H}_0$ and $\mathbf{q} \cong q \, \left\{ 0, \sin\theta, \cos\theta\right\}$ (see Fig.~\ref{fig1}); the atomic magnetic form factor (in the expression for $b_H$) is approximated to unity since we are dealing with forward scattering. 

Equation~(\ref{sigmasansperpunpol}) can be separated into the nuclear and magnetic SANS cross section at saturation, the so-called residual SANS cross section $d \Sigma_{\mathrm{res}} / d \Omega$, and into the spin-misalignment SANS cross section $d \Sigma_M / d \Omega$, \textit{i.e.},
\begin{equation}
\label{sigmasansperp2d}
\frac{d \Sigma}{d \Omega}(\mathbf{q}) = \frac{d \Sigma_{\mathrm{res}}}{d \Omega}(\mathbf{q}) + \frac{d \Sigma_M}{d \Omega}(\mathbf{q}) ,
\end{equation}
where
\begin{equation}
\label{sigmaresperp}
\frac{d \Sigma_{\mathrm{res}}}{d \Omega}(\mathbf{q}) = \frac{8 \pi^3}{V} \left( |\widetilde{N}|^2
 + b_H^2 |\widetilde{M}_s|^2 \sin^2\theta \right)
\end{equation}
and
\begin{eqnarray}
\label{sigmasmperp}
\frac{d \Sigma_M}{d \Omega}(\mathbf{q}) = \frac{8 \pi^3}{V} b_H^2 \left( |\widetilde{M}_x|^2 + |\widetilde{M}_y|^2 \cos^2\theta \right. \nonumber \\ \left. - (\widetilde{M}_y \widetilde{M}_z^{\ast} + \widetilde{M}_y^{\ast} \widetilde{M}_z) \sin\theta \cos\theta \right) .
\end{eqnarray}
The residual SANS cross section contains the nuclear scattering and the magnetic SANS due to nanoscale spatial variations of the saturation magnetization ($\propto |\widetilde{M}_s(\mathbf{q})|^2$) [compare Eqs.~(\ref{msatdef} and (\ref{mvecdeffourier})], whereas the spin-misalignment SANS cross section contains the magnetic SANS due to spatial variations in the \textit{orientation and magnitude} of the magnetization. Note that in writing down Eq.~(\ref{sigmasmperp}) we have made the approximation that $|\widetilde{M}_s|^2 \cong |\widetilde{M}_z|^2$, which may be justified in the approach-to-saturation regime.

\begin{figure}[tb]
\centering
\resizebox{1.0\columnwidth}{!}{\includegraphics{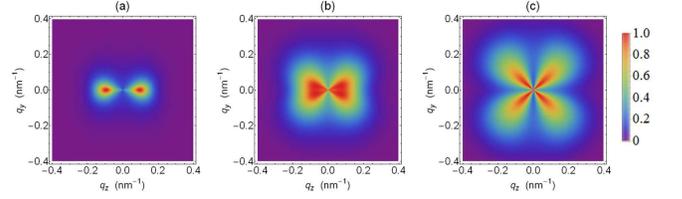}}
\caption{Contour plots of $d \Sigma_M / d \Omega$ at selected applied magnetic fields [Eq.~(\ref{sigmasmperp})] ($\mathbf{k}_0 \perp \mathbf{H}_0$). $H_i$ values (in T) from left to right: 0.4; 0.8; 4.2. All data were normalized to unity by the respective maximum value.}
\label{fig4}
\end{figure}
\begin{figure}[tb]
\centering{\includegraphics[width=0.70\columnwidth]{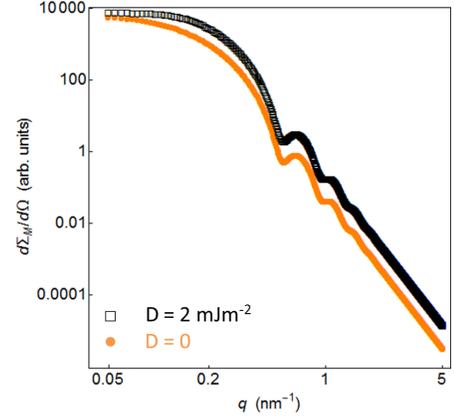}}
\caption{\label{fig5} Azimuthally-averaged $d \Sigma_M / d \Omega$ at $\mu_0 H_i = 0.8 \, \mathrm{T}$ with and without the DM term (see inset) (log-log scale) ($\mathbf{k}_0 \perp \mathbf{H}_0$). Both $d \Sigma_M / d \Omega$ have been convoluted with the same Gaussian distribution function ($\overline{R} = 8 \, \mathrm{nm}$; $\sigma = 0.7$) for both $\widetilde{H}^2_p$ and $\widetilde{M}^2_z$.}
\end{figure}

The spin-misalignment SANS cross section $d \Sigma_M / d \Omega$ is shown in Fig.~\ref{fig4}. With increasing field, $d \Sigma_M / d \Omega$ changes its angular anisotropy from an elliptically-distorted pattern with maxima along the horizontal direction [Fig.~\ref{fig4}(a)] to a clover-leaf-type anisotropy [Fig.~\ref{fig4}(b) and \ref{fig4}(c)]. Figure~\ref{fig5} displays the (over $2\pi$) azimuthally-averaged $d \Sigma_M / d \Omega$ at $\mu_0 H_i = 0.8 \, \mathrm{T}$ with and without the DM term. Since the DM interaction introduces nonuniformity into the spin structure, the spin-misalignment scattering cross section is larger when this term is included.

\subsection{Spin-flip SANS ($\mathbf{k}_0 \perp \mathbf{H}_0$)}

\begin{figure*}[tb]
\centering{\includegraphics[width=1.0\textwidth]{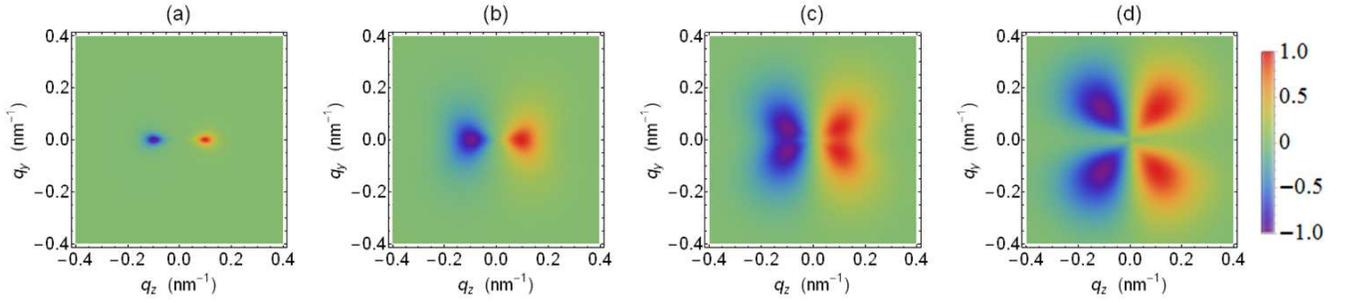}}
\caption{\label{fig6} Contour plots of the spin-flip difference cross section $-2i \chi(\mathbf{q})$ [Eq.~(\ref{ffinal})] at selected applied magnetic fields ($\mathbf{k}_0 \perp \mathbf{H}_0$). $H_i$ values (in T) from left to right: 0.3; 0.6; 1.0; 3.5. All data were normalized to unity by the respective maximum value.}
\end{figure*}

Assuming a perfect neutron optics and neglecting nuclear spin-incoherent SANS, the spin-flip SANS cross section of a bulk ferromagnet can be written as: \cite{maleyev63,blume63}
\begin{eqnarray}
\label{sigmasansperpsf}
\frac{d \Sigma^{\pm \mp}}{d \Omega}(\mathbf{q}) = \frac{8 \pi^3}{V} \, b_H^2 \left( |\widetilde{M}_x|^2 + |\widetilde{M}_y|^2 \cos^4\theta \right. \nonumber \\ \left. + |\widetilde{M}_z|^2 \sin^2\theta \cos^2\theta \right. \nonumber \\ \left. - (\widetilde{M}_y \widetilde{M}_z^{\ast} + \widetilde{M}_y^{\ast} \widetilde{M}_z) \sin\theta \cos^3\theta \mp i \chi(\mathbf{q}) \right) , 
\end{eqnarray}
where 
\begin{eqnarray}
\label{sfcross}
\chi(\mathbf{q}) & = & \left( \widetilde{M}_x \widetilde{M}_y^{\ast} - \widetilde{M}_x^{\ast} \widetilde{M}_y \right) \cos^2\theta   \nonumber \\ & & - \left( \widetilde{M}_x \widetilde{M}_z^{\ast} - \widetilde{M}_x^{\ast} \widetilde{M}_z \right) \sin\theta \cos\theta .
\end{eqnarray}
We remind the reader that $\widetilde{M}_z$ is assumed to be real-valued and isotropic. The first superscript (\textit{e.g.}, ``$+$'') that is attached to $d \Sigma / d \Omega$ in Eq.~(\ref{sigmasansperpsf}) refers to the spin state of the incident neutrons, whereas the second one (\textit{e.g.}, ``$-$'') specifies the spin state of the scattered neutrons. Obviously, $\chi = 0$ when $\widetilde{M}_{x, y, z}$ are real-valued ($l_D = 0$). Inserting the expressions for the Fourier coefficients [Eqs.~(\ref{solmxqx0}) and (\ref{solmyqx0})] and averaging over the direction of the magnetic anisotropy field yields for the difference cross section $- 2 i \chi(\mathbf{q}) = \frac{d \Sigma^{+-}}{d \Omega} - \frac{d \Sigma^{-+}}{d \Omega}$:
\begin{widetext}
\begin{equation}
\label{ffinal}
- 2 i \chi(\mathbf{q}) = \frac{2 \widetilde{H}^2_p p^3 \left( 2 + p \sin^2\theta \right) l_D q \cos^3\theta + 4 \widetilde{M}_z^2 p (1 + p)^2 l_D q \sin^2\theta \cos\theta}{\left( 1 + p \sin^2\theta - p^2 l_D^2 q^2 \cos^2\theta \right)^2} .
\end{equation}
\end{widetext}
Using the magnetic and structural parameters of Table~\ref{table1} as well as the sphere form factor for both the anisotropy-field Fourier coefficient $\widetilde{H}^2_p(q R)$ and for the longitudinal magnetization Fourier coefficient $\widetilde{M}^2_z(q R)$, the quantity $-2i \chi(\mathbf{q})$ is plotted in Fig.~\ref{fig6}.

\begin{figure}[tb]
\centering{\includegraphics[width=0.90\columnwidth]{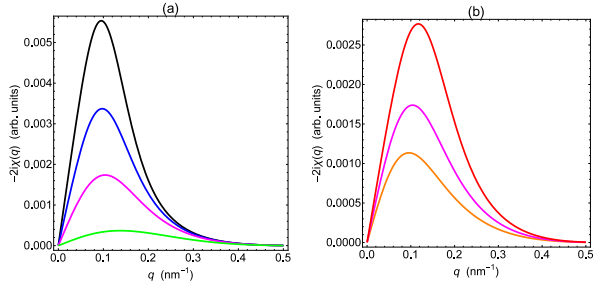}}
\caption{\label{fig7} Azimuthal average of the spin-flip difference cross section, $-2i \chi(q, H) = -2i \int_0^{\pi/2} \chi(q, H, \theta) d\theta$, at (a) selected applied magnetic fields and constant $D = 2.0 \, \mathrm{mJ/m^2}$, and (b) for constant field $\mu_0 H_i = 0.8 \, \mathrm{T}$ but varying DM constant $D$ ($\mathbf{k}_0 \perp \mathbf{H}_0$). The field (in T) in (a) increases from top to bottom: 0.5; 0.6; 0.8; 2.0. The DM constant (in $\mathrm{mJ/m^2}$) in (b) increases from bottom to top: 1.5; 2.0; 2.5.}
\end{figure}

At small fields, the asymmetry of the pattern is such that two extrema parallel and antiparallel to the field axis are observed [Fig.~\ref{fig6}(a) and \ref{fig6}(b)], whereas at larger fields additional maxima and minima appear approximately along the detector diagonals [Fig.~\ref{fig6}(c) and \ref{fig6}(d)]. The azimuthally-averaged function $-2i \chi(q, H)$ is shown in Fig.~\ref{fig7} as a function of the applied magnetic field $H_i$ for a fixed value of the DM constant $D$ [Fig.~\ref{fig7}(a)] and as a function of $D$ for a fixed $H_i$-value [Fig.~\ref{fig7}(b)]. The strong field dependency of $-2i \chi(q, H)$ may be employed in order to experimentally determine the DM constant. 

For statistically isotropic systems (\textit{e.g.}, polycrystalline magnetic materials), the predicted effect may not be observable due to the random orientation of the interfaces (grain boundaries) and the ensuing random orientation of the DM vectors. Therefore, one strategy to observe the polarization dependence of the spin-flip SANS cross section might be experiments on heavily deformed (cold-worked) magnets possessing a texture axis \cite{Fedorov1997} or on field-cooled nanocrystalline rare-earth magnets. \cite{lott08} For the latter, nanocrystallinity is required in order to guarantee a high defect (interface) density, whereas field cooling from the paramagnetic state at room temperature to a low-temperature ferromagnetic state may orient the DM vectors on the interfaces.

\section{Conclusion}

Within the framework of the continuum theory of micromagnetics, we have investigated the influence of the Dzyaloshinski-Moriya (DM) interaction on the elastic magnetic small-angle neutron scattering (SANS) cross section of bulk ferromagnets. Due to the complex character of the magnetization Fourier components, a polarization dependence of the spin-flip SANS cross section is predicted [Eq.~(\ref{ffinal})]. This effect may be experimentally studied on (field-cooled) nanocrystalline rare-earth magnets or on heavily cold-worked magnetic materials, which may provide a means to further scrutinize the DM interaction.

\section*{Acknowledgements}

We thank the National Research Fund of Luxembourg for financial support (Project No.~INTER/DFG/12/07).

\bibliographystyle{apsrev4-1}
\end{document}